\documentclass[letterpaper, 10 pt, conference]{IEEEtran} 

\usepackage{hhline}
\usepackage{color}
\usepackage{amsmath}
\usepackage{graphicx}
\usepackage{fixmath}
\usepackage[font=small,labelfont=bf,
   justification=justified,
   format=plain]{caption} 
\usepackage{url}
\usepackage{array}
\usepackage{makecell}
\usepackage{adjustbox}
\usepackage{physics}
\usepackage[dvipsnames]{xcolor}
\usepackage{amssymb}
\usepackage{soul}
\usepackage{caption}
\usepackage{subcaption}

\renewcommand\IEEEkeywordsname{Keywords}
{

\title{\Huge \bf
Comparison of different CNNs for breast tumor classification from ultrasound images
}

\author{Jorge F. Lazo$^{1}$, Sara Moccia$^{2,3}$, Emanuele Frontoni$^{3}$ and Elena De Momi$^{1}$ \\
{\textit{$^{1}$	Department of Electronics, Information and Bioengineering}, Politecnico di Milano, Milan, Italy}\\
\textit{$^{2}$	Department of Advanced Robotics, Istituto Italiano di Tecnologia, Genoa, Italy}\\
\textit{$^{3}$ Department of Information Engineering, Università Politecnica delle Marche, Ancona, Italy}
}

\begin{document}

\maketitle
\thispagestyle{empty}
\pagestyle{empty}

\begin{abstract}
Breast cancer is one of the deadliest cancer worldwide. A timely detection could reduce mortality rates. In the clinical routine, classifying benign and malignant tumors from ultrasound (US) imaging is a crucial but challenging task. An automated method, which can deal with the variability of data is therefore needed. 
In this paper, we compared different Convolutional Neural Networks (CNNs) and transfer learning methods for the task of automated breast tumor classification. The architectures investigated in this study were VGG-16 and Inception V3. Two different training strategies were investigated: the first one was using pretrained models as feature extractors and the second one was to fine tune the pretrained models. A total of 947 images were used, 587 corresponded to US images of benign tumors and 360 with malignant tumors. 678 images were used for the training and validation process, while 269 images were used for testing the models.  
Accuracy and Area Under the receiver operation characteristic Curve (AUC) were used as performance metrics. The best performance was obtained by fine tuning VGG-16, with an accuracy of 0.919 and an AUC of 0.934. The obtained results open the opportunity to further investigation with a view of improving cancer detection.

\end{abstract}


\section{INTRODUCTION}
\label{sec:intro}

Breast cancer is the most common cancer in women \cite{cancer_2019}. Cancer screening is performed via Breast Ultrasound (BUS) imaging and mammography. 
BUS is recommended in a large variety of cases, such as women under the age of 30 and/or in case of pregnancy.  
Cancer diagnosis is performed in the clinical practice by clinicians through visual BUS-image analysis. However, it is well known that BUS image acquisition and analysis are highly dependent on the clinician level of expertise \cite{widely_variability}. 


Computer-aided diagnosis (CAD) systems for BUS image analysis have recently shown to be able to tackle the variability associated with both breast anatomy and BUS images, becoming a suitable tool to improve diagnosis accuracy \cite{XIAN2018340}.


In the last few years, deep-learning (DL) approaches, and more specifically convolutional neural networks (CNNs), have become the standard in research for BUS-image analysis \cite{LIU2019261}. However, there are still open challenges that need to be addressed. 
Among them, the necessity of relying on a large and annotated BUS dataset for CNN training \cite{LIU2019261}. 
A possible solution to attenuate this issue could be to exploit transfer learning and fine tuning. These kind of approaches has already been applied in other studies and imaging modalities presenting promising results as reported in \cite{trasnfer_deep,fiorentino2019learning,calimeri2016optic}. However, even transfer learning seems to be the way of proceeding, there are different techniques to perform this method, and to our knowledge, there is no study in the comparison of this strategies applied to this kind of data. 
%

In this paper we explore  the use of pre-trained existing models and two different training strategies with the aim of determining which of these strategies and model is more suitable for the task of breast tumor classification. 
The paper is organized as follows: Sec.~\ref{sec:soa} reviews the state of the art in automated tumor classification methods, Sec.~\ref{sec:meth} delve into the methodology explaining the architectures and transfer learning techniques used. In Sec.~\ref{sec:res} information about the dataset used and details about the training is provided. 
In Sec.~\ref{sec:exp_results} the results are presented and discussed. Finally, Sec.~\ref{sec:concl}, concludes this paper discussing results and proposing future improvements. 



\section{Deep Learning in BUS image analysis}
\label{sec:soa}

In the last few years, DL methods and specifically CNNs have become the state of the art for image analysis tasks. 
The application of DL in the analysis of medical US images involves different specific tasks, such as classification, segmentation, detection, registration, as well as the development of new methodologies for image-guided interventions. 

In the specific case of tumor classification, different extensions and variations of DL approaches have been developed. In \cite{yap2017automated} the authors propose the use of different CNNs for locating regions of interest (ROIs) corresponding to lesions.  
In \cite{bakkouri2017breast}, CNNs are used as feature extractors and the features obtained are classified with a Support Vector Machine (SVM).
%
%
In \cite{Bashir_BUS} an architecture based on AlexNet is proposed, and its performance is compared with some pretrained models, using a small custom-built dataset. In \cite{al2019deep},  a method based on the use of Generative Adversarial Networks (GANs) for data augmentation is proposed, later the authors compare the performance of this network in the task of generating synthetic data, using pretrained models, in this case VGG16, Inception, ResNet and NasNet. A further step is taken in \cite{MOON2020105361}, where the authors propose the use of ensembles to develop a better and more comprehensive generalized model. Their model is based in the use of VGG16-like architectures as well as different versions of ResNet and DenseNet. In  \cite{han2017deep} a modification to the GoogLeNet architecture is proposed, this network is an early version of Inception V3 and it is composed by a main branch and two auxiliary classifiers, specifically they suggest to remove the auxiliary classifiers from the main branch of the network. In \cite{xiao2018comparison} they compare different pre-trained models but only using fine-tuning. In this study they compare the CNNs ResNet50, InceptionV3, and Xception.

\section{PROPOSED METHODS}
\label{sec:meth}

In this work, we investigated two different transfer learning techniques: (i) fine tuning and (ii) using pretrained CNNs as feature extractor. Each of these methods were tested using two different CNN architectures:

\subsection{CNN architectures}

\begin{figure}[tbp]
        \centering
        \includegraphics[width=0.45\textwidth, height=5cm]{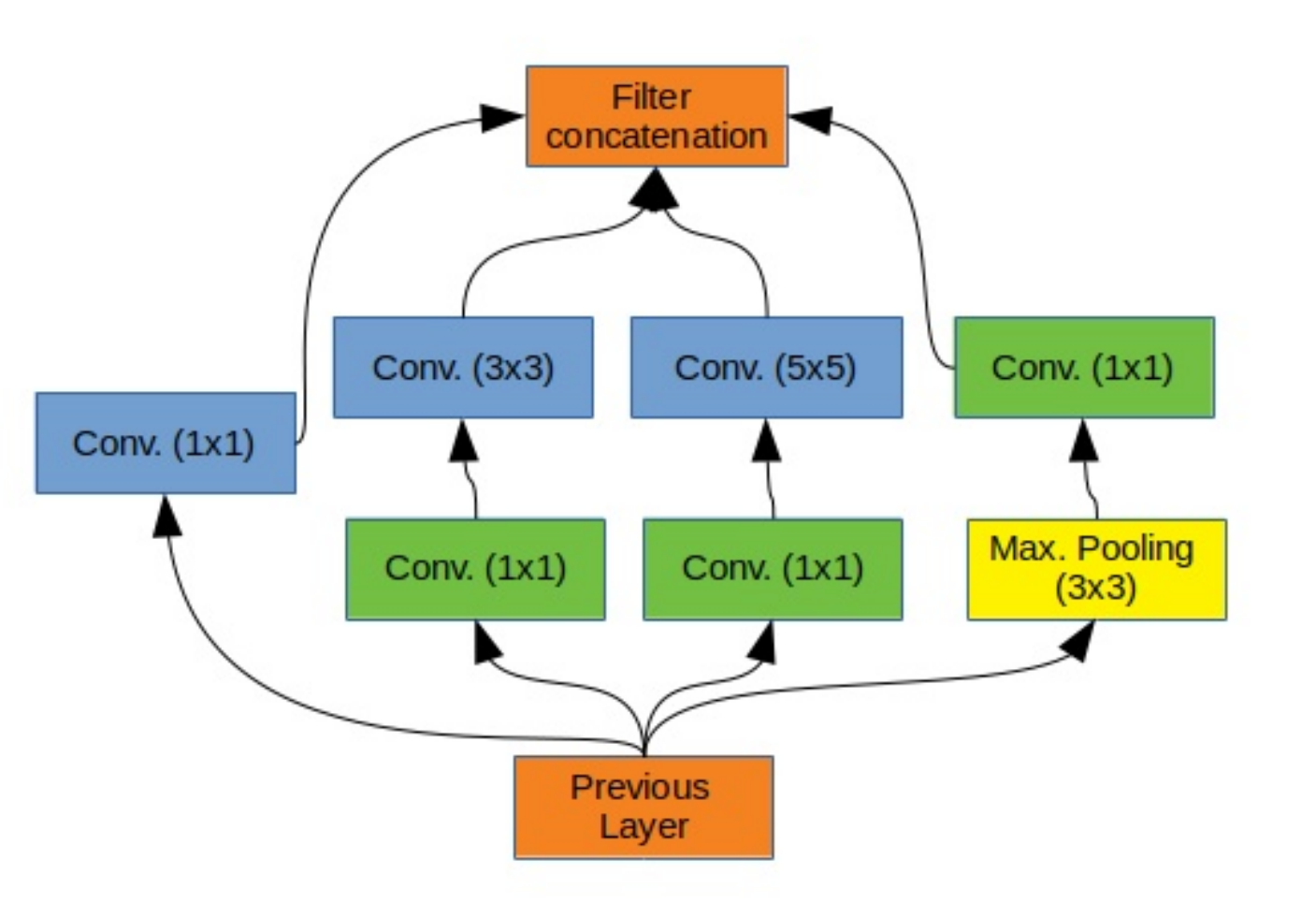}
        \caption{Diagram of the inception module. The module is composed by several convolutional kernels of size 1x1, 3x3 and 5x5 connected in parallel. The 1x1 kernels placed before the 3x3 and 5x5 is used to reduce the dimensionality of the feature map. The output of each of the branches is finally concatenated before getting into the next stage.}
        \label{fig:inception_module}
    \end{figure}

\begin{itemize}

    \item  VGG-16 is a 16-layer CNN model which has a sequential architecture consisting of 13 convolutional layers and 5 max-pooling layers \cite{vgg_16_paper}. The architecture starts with a convolutional layer with 64 kernels. This number is doubled after each pooling operation until it reaches 512. The pooling layer is placed after selected convolutional layers in order to reduce dimension in the activation maps and hence  of the subsequent convolution layers. This in general reduces the number of parameters that the CNN needs to learn. The convolutional kernel size of all the convolutional layers in this model is 3x3. The model ends with three fully-connected (FC) layers with 64 neurons each, which perform the classification. 

    \item  Inception V3 \cite{InceptionV3} uses an architectural block called inception module, which consists of convolutional kernels with different sizes (1x1, 3x3 and 5x5) that are connected in parallel (Fig. \ref{fig:inception_module}). The use of different kernel sizes allows the identification of image features at different scales. 
    %
    Furthermore, Inception V3 uses not just one classifier but two, the second one is an auxiliary classifier which is used as regularizer. 
    %
    One of the main advantages of this model is that it is composed of about 23 million of parameters even it has 42 layers, therefore the computational cost for training this network is also less than the one needed to retrain VGG-16. However, given its more complex topology 
    it is also harder to retrain. 
\end{itemize}

\begin{figure}[tbp]
    \centering
    \includegraphics[height= 2.5cm, width=0.2\textwidth]{images/benign.pdf}
    \includegraphics[height= 2.5cm, width=0.2\textwidth]{images/malignant.pdf}
    \caption{Samples of two images from the BUSI dataset. Image with a benign tumor (left) and with a malignant tumor (right).}
    \label{fig:tumor_images}
\end{figure}

\subsection{Transfer learning}

As introduced in Sec. \ref{sec:intro}, training a CNN model from scratch requires a large number of computational resources as well as a fair amount of labeled data. It also often requires a considerable amount of time, even using several graphics processing units (GPUs). 

Transfer learning allows making the training process more efficient by using a model that has already been trained on a different dataset. There are different ways to perform transfer learning, in this work fine tuning and feature extraction techniques were explored. The first one refers to re-adjust the weights of the new distribution of the new training data, i.e. to tune the weights of the CNN by training it on the new dataset for few epochs and with a low learning rate. 
Usually, only the weights in the last layers are retrained. 
In this work, both VGG 16 and Inception V3 were pretrained on  ImageNet\footnote{\url{http://www.image-net.org/}}, a dataset which has more than 14 million images belonging to 1000 classes. The second method, feature extraction, refers to use the whole network as a feature identifier, and then making use of the high level features, obtained from the network, in another classifier.

The fine-tuning strategy was performed by replacing the last fully connected layer composed by 2 neurons (one for the benign and the other for the malignant class), and then fine-tune different number of layers in the network. The layers chosen to be fine-tuned were  the very last one up to the last 3 layers of each network. 

In the case of the feature extraction method, an additional classifier was added and trained, this was composed by a Global Average Pooling Layer, a Fully connected layer with a Rectified Linear Unit (\emph{ReLU}) activation function. Finally, a fully connected layer with \emph{softmax} activation function and 2 neurons. 



\section{RESULTS}
\label{sec:exp_results}

\subsection{Dataset}
The data used for this work came from two public available datasets collected by the groups of Rodriguez et al.\footnote{https://data.mendeley.com/datasets/wmy84gzngw/1} and Fahmy et al.\footnote{https://scholar.cu.edu.eg/?q=afahmy/pages/dataset}  
The first dataset consists of 250 breast tumor images (100 benign and 150 malignant) with an average size of 100x75 pixels.    
The second one consists of 963 images of an average image size of 500x500 pixels, in this dataset 487 images correspond to images with benign tumors, 210 with malignant tumors and 266 images with any tumor at all. For this project only the images with benign and malignant tumors were used. 

Hence, the dataset used in this work consisted of 537 and 360 with benign and malignant tumors, respectively. The whole dataset was shuffled randomly, and then split, in a stratified fashion, in two subsets, with 630 images for training and validation, and 269 for testing. A sample of benign and malignant BUS images is shown in Fig.~\ref{fig:tumor_images}. 

During the training of the networks, mini-batch gradient descent method was applied and Adam optimization algorithm. At the moment of creating the training batches the images were reshaped \textcolor{blue}{}``on the fly'' to match the size of the input of each network, 224x224 pixels in the case of VGG-16 and 299x299 pixels for the case of Inception V3. 

\subsection{Training settings}

Training was performed with an initial learning rate of 0.001. The size of the output for the added fully connected (FC) layer was 1024 and 512 for Inception V3 and VGG-16 respectively. The mini-batch size used was of 50 images. 


The experiments were carried out on an Nvidia GPU GTX 1660 using Keras with TensorFlow backend.

\subsection{Performance Metrics}

To evaluate the proposed CNNs, the area (AUC) under the Receiver Operating Characteristic (ROC) curve was computed.
The accuracy ($ACC$), was computed as follows:



\begin{equation}
ACC = \frac{TP+TN}{TP+FP+FN+TN} 
\end{equation}
where $TP$ and $TN$ are the amount of malignant BUS images correctly classified, respectively, and $FN$ and $FP$ are the amount of malignant BUS images missclassified. 



\begin{figure}[tbp]
    \begin{subfigure}[b]{0.5\textwidth}
        \centering
    \includegraphics[width=0.79\textwidth, height = 5.8cm]{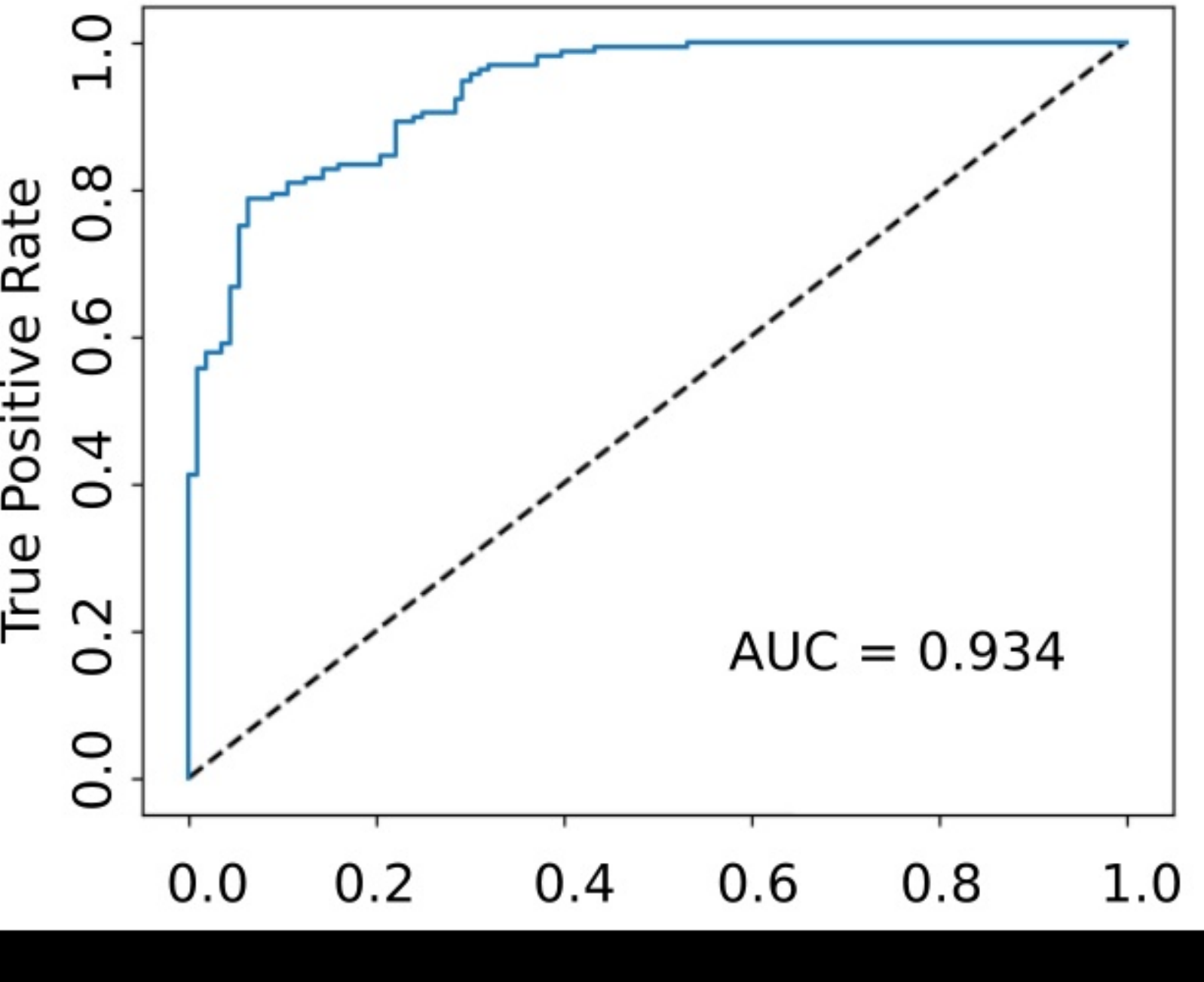}
    \caption{}
    \end{subfigure}
    
    \hspace{0.25cm}
    \begin{subfigure}[b]{0.5\textwidth}
        \centering
    \includegraphics[width=0.9\textwidth, height =6.1cm]{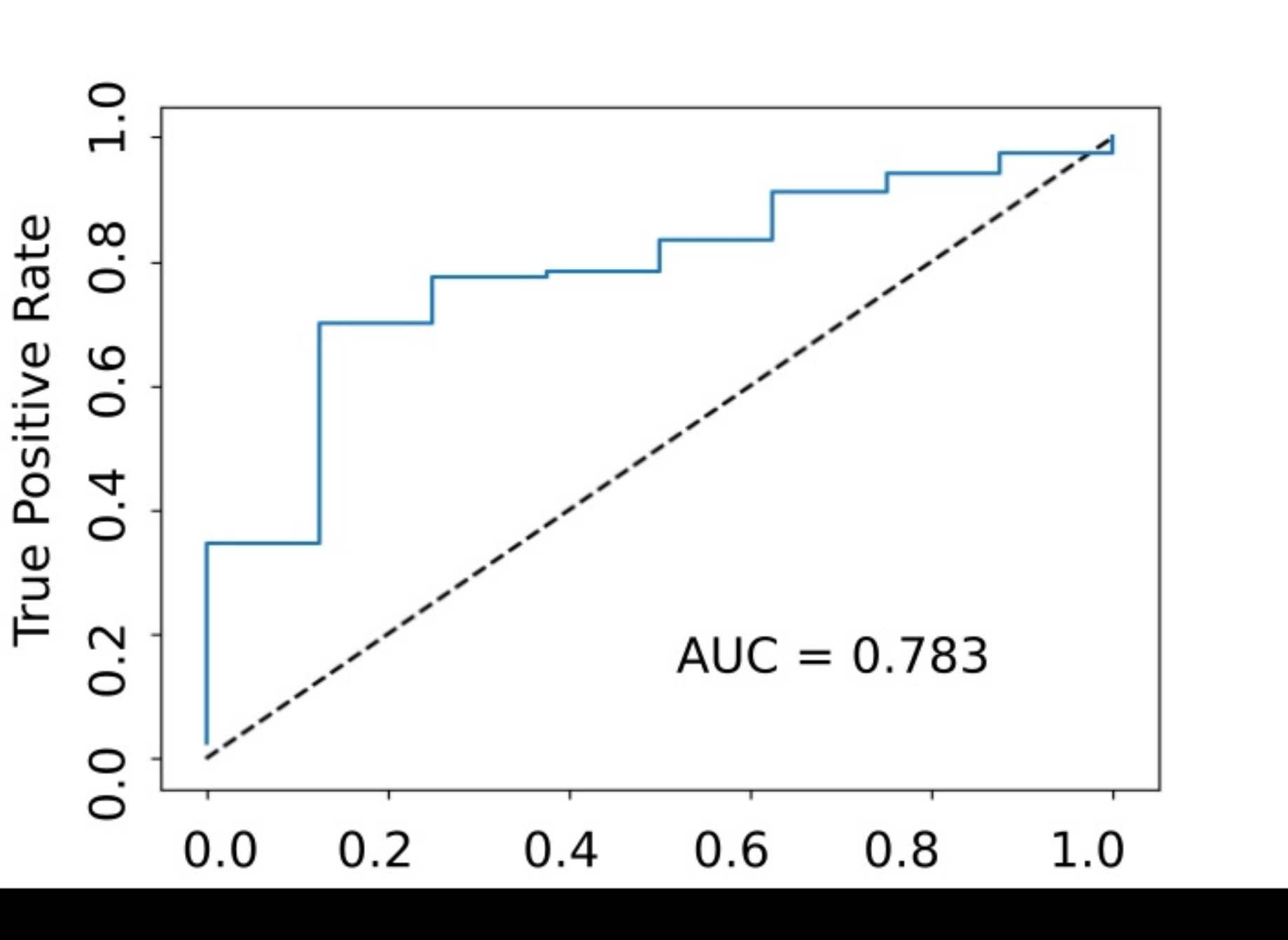} 
    \caption{}
    \end{subfigure}
    \caption{Receiver operating characteristics (ROC) curves for the 2 different models tested using fine-tuning: (a) VGG-16, (b) Inception V3. }
    \label{fig:ROC_curves_result}
\end{figure}

\begin{table}[tbp]
    \centering
    \caption{Comparison of the obtained results with VGG-16 and Inception V3 with different training methods each one. The numbers highlighted in black correspond to the model and method that obtained the highest scores in terms of \emph{ACC} and \emph{AUC}.}
    \begin{tabular}{c|c|c|c}
        \hline
        Model & trainable parameters & $ACC$ & $AUC$ \\ \hline
        \multicolumn{4}{c}{Feature extraction} \\ \hline \hline
        VGG-16 & 512,512 & 0.862 & 0.791  \\ \hline
        Inception V3 & 1,311,744 & 0.713 & 0.623 \\ \hline
        \multicolumn{4}{c}{Fine-tuning} \\ \hline \hline
        VGG-16 & 1,054,722 & \textbf{0.919} & \textbf{0.934}  \\ \hline
        Inception V3 & 2,388,539 & 0.756 & 0.783  \\ \hline
    \end{tabular}
    \label{tab:classification_results}
\end{table}


\section{RESULTS AND DISCUSSION}
\label{sec:res}

The obtained ROC curves for VGG-16 and Inception V3 using the fine-tuning are shown in Fig. \ref{fig:ROC_curves_result}, a summary of the results obtained with each network and each training method is presented in Table \ref{tab:classification_results}. Fine tuning worked better, and the model which performed better was VGG-16 in both cases. 
However, the difference between these two tranfer learning methods is smaller with Inception V3. In the case of the ACC, the difference is only of 0.046 while the difference of the AUC is 0.16. Using fine tuning on VGG-16, the values of ACC = 0.919 and AUC = 0.934 were obtained. 

\begin{figure}[tbp]
    \centering
    \begin{subfigure}[b]{0.15\textwidth}
         \includegraphics[width = 2cm, height = 2cm]{images/benign-100.pdf}
         \caption{}
     \end{subfigure}
     \begin{subfigure}[b]{0.15\textwidth}
         \includegraphics[width = 2cm, height = 2cm]{images/benign-131.pdf}
         \caption{}
     \end{subfigure}
     \begin{subfigure}[b]{0.15\textwidth}
         \includegraphics[width = 2cm, height = 2cm]{images/benign-207.pdf}
         \caption{}
     \end{subfigure}
    \begin{subfigure}[b]{0.15\textwidth}
         \includegraphics[width = 2cm, height = 2cm]{images/benign-196.pdf}
         \caption{}
     \end{subfigure}
     \begin{subfigure}[b]{0.15\textwidth}
         \includegraphics[width = 2cm, height = 2cm]{images/benign-281.pdf}
         \caption{}
     \end{subfigure}
     \begin{subfigure}[b]{0.15\textwidth}
         \includegraphics[width = 2cm, height = 2cm]{images/benign-355.pdf}
         \caption{}
     \end{subfigure}
    \begin{subfigure}[b]{0.15\textwidth}
         \includegraphics[width = 2cm, height = 2cm]{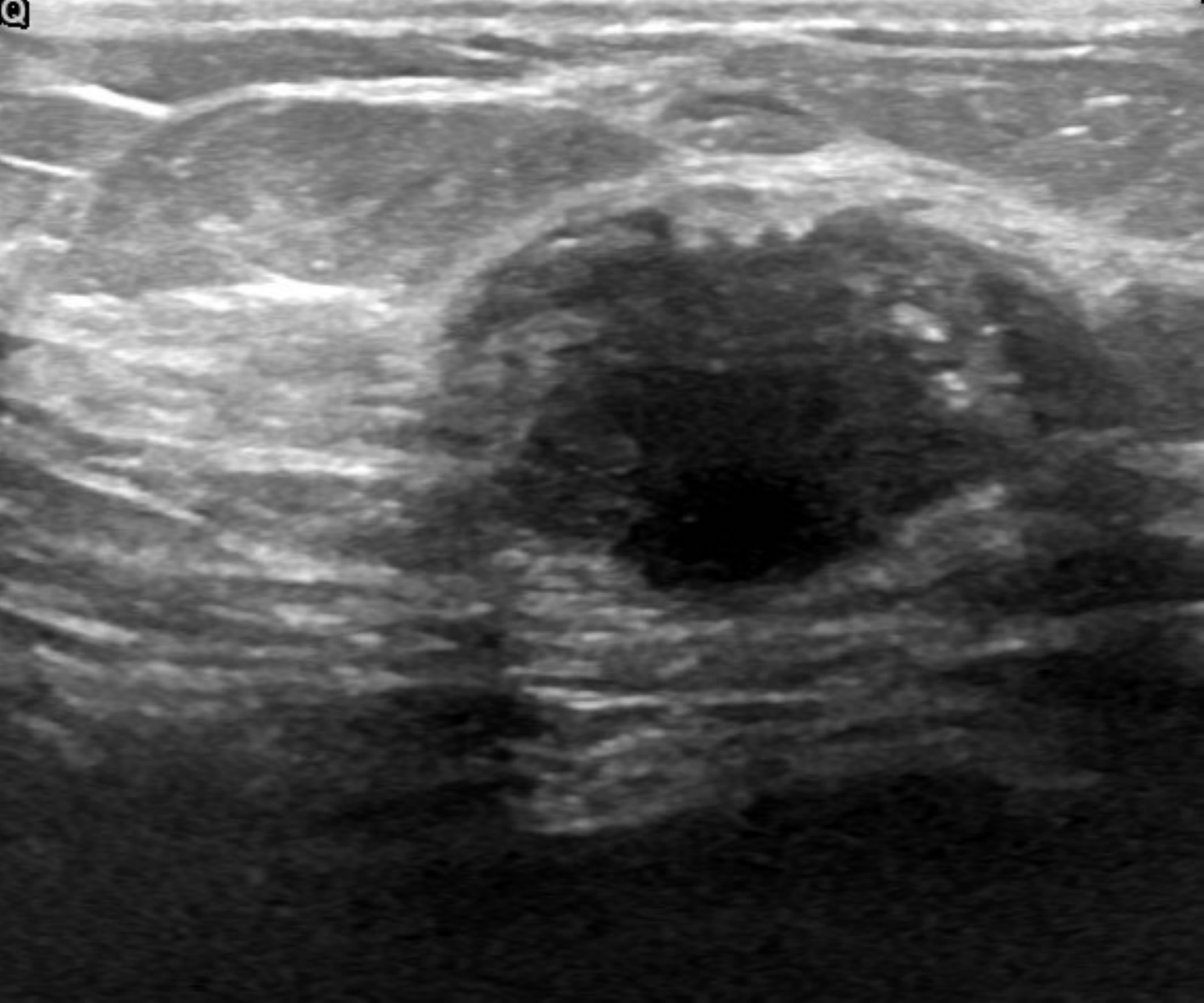}
         \caption{}
     \end{subfigure}
     \begin{subfigure}[b]{0.15\textwidth}
         \includegraphics[width = 2cm, height = 2cm]{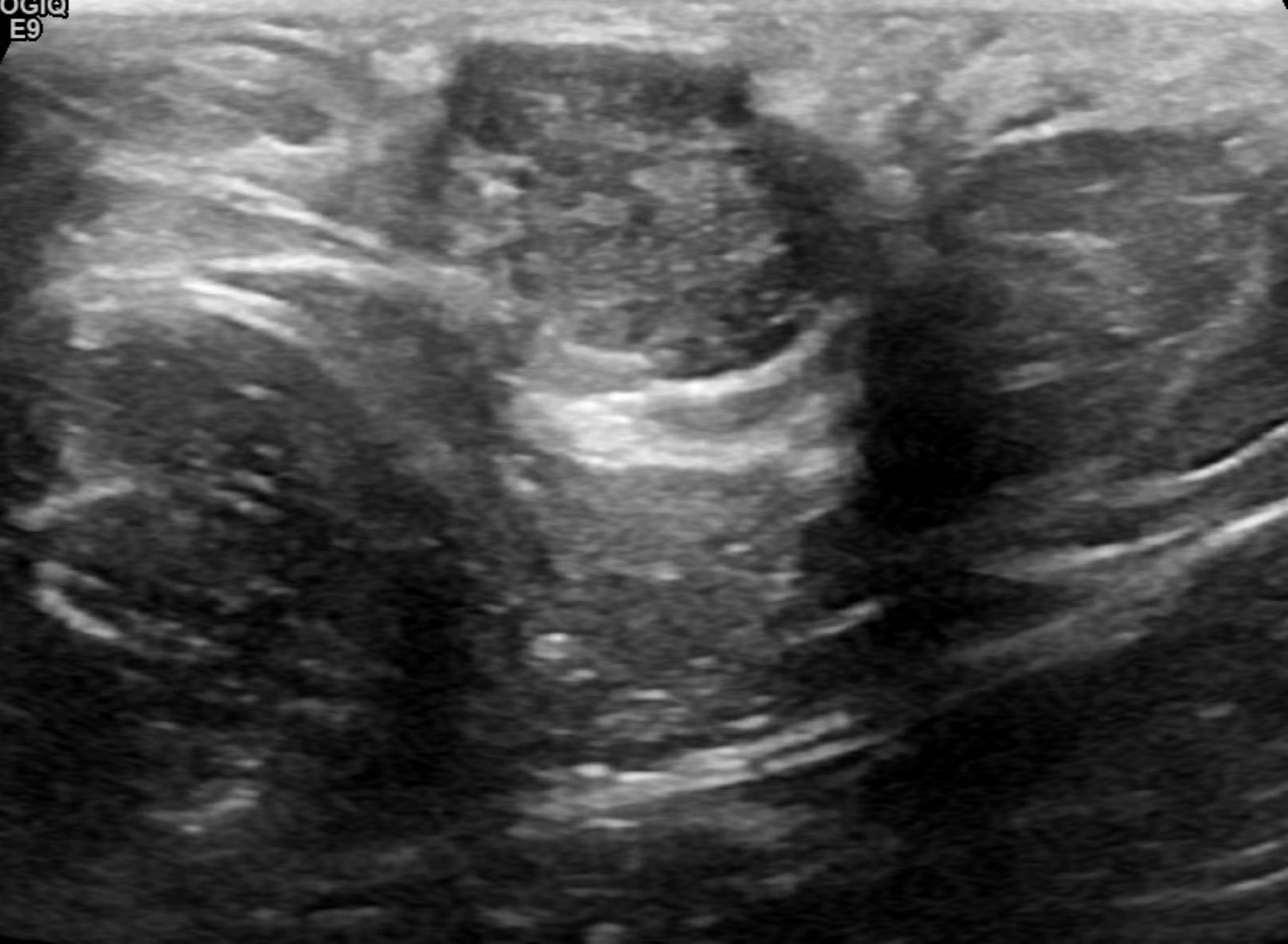}
         \caption{}
     \end{subfigure}
     \begin{subfigure}[b]{0.15\textwidth}
         \includegraphics[width = 2cm, height = 2cm]{images/malignant_154_.pdf}
         \caption{}
     \end{subfigure}
    \begin{subfigure}[b]{0.15\textwidth}
         \includegraphics[width = 2cm, height = 2cm]{images/malignant_160_.pdf}
         \caption{}
     \end{subfigure}
     \begin{subfigure}[b]{0.15\textwidth}
         \includegraphics[width = 2cm, height = 2cm]{images/malignant_203_.pdf}
         \caption{}
     \end{subfigure}
     \begin{subfigure}[b]{0.15\textwidth}
         \includegraphics[width = 2cm, height = 2cm]{images/malignant_9_.pdf}
         \caption{}
     \end{subfigure}
    \caption{Sample of results obtained by using fine tuning and VGG-16 architecture. (a)-(c) images with benign tumors missclassified as malign, (d)-(f) images with benign tumors correctly classified, (g)-(i) images with malign tumors missclassified as benign, (j)-(l) images with malign tumors correctly classified.}
    \label{fig:sample_results}
\end{figure}

Inception V3 only reaches the values of 0.756 and 0.783, respectively in the test dataset. However, during the training, it reaches accuracy values over 0.93 and AUC of 0.89 which implies that the model is over-fitting. Normalization techniques, such as Dropout and $L_{2}$ normalization, may help to reduce the over-fitting issue and improve its performance in the testing stage \cite{Phaisangittisagul}.  
A deep exploration of the hyperparameters choice such as the batch size, the learning rate, the number of layers to be retrained (for the case of fine tuning), the size of the fully connected layers and number of layers (for the case of feature extraction) and the use of normalization methods is needed to be carried out. As is it possible to see from the sample of results show in Fig. \ref{fig:sample_results} the variability of the content in the images affects the predictions made by the network. 
This may could be tackled by using different data augmentation methods which can generalize the variability of this kind of data. 




\section{CONCLUSION}
\label{sec:concl}

In this paper, we explored the application of  CNN for the task of tumor classification using BUS images. We trained two pretrained models in two different ways: using the existing models as feature extractors and fine tuning them. We observed that for this case fine tuning achieved better results and in general the architecture VGG-16 performed better in the test dataset. The best value obtained for accuracy was 0.919 with an AUC of 0.934. In the case of Inception V3 the values obtained were considerably lower, 0.756 and 0.783 respectively, but its performance during the training was better, reaching an accuracy value of 0.93. The results obtained using fine tuning are in the same range as the one of the state of the art, but we can state that fine tuning is the strategy to go and which should be developed further. Further normalization methods will be needed in order to avoid over-fitting as it seems to be the case. Future work includes the use of data augmentation, the testing of different and novel architectures which have claimed to be more robust against perspective transformations, spatial orientation, scale changes, etc. such as Capsule Networks and the exploration of image segmentation with other architectures. 

\section*{Acknowledgements}

This project has received funding from the European Union’s Horizon 2020 research and innovation programme under the Marie Sklodowska-Curie grant agreement No 813782. 

\bibliographystyle{IEEEtran}
\bibliography{biblio} 
\end{document}